\documentclass[12pt]{article}
\usepackage{graphicx}
\usepackage{mathrsfs, amssymb, eucal, amsmath, amsthm, amscd, amsrefs}

\numberwithin{equation}{section}

\theoremstyle{remark}

\begin{document}

\vspace{7mm}

\centerline{\bf AN ADHESION MODEL FOR THE DRAG FORCE} \vspace{2mm}

\centerline{ Dan Com\u{a}nescu}

 \vspace{4mm}

\footnotesize{\textit{West University of Timi\c{s}oara, Faculty of Mathematics and Computer Science, Department of Mathematics, Bd. V.
P\^{a}rvan no. 4, Timi\c{s}oara, Romania, e-mail: comanescu@math.uvt.ro}}

\vspace{4mm}

{\footnotesize

 \noindent {\bf Abstract.} The paper present a model for the drag force between a resistive medium and a solid body using the hypothesis that
the drag force is created by the adhesion of some particles of the resistive medium on the solid body's surface. The study focus on the mass
evolution of solid body.



}

\section{Introduction}

\hspace{0.5cm} A drag force usually models the influence of a resistive medium on a solid body. If the body is a particle then the drag force
has the form:
\begin{equation}\label{*}
\overrightarrow{F_{f}}=-\varphi (v)\overrightarrow{v}
\end{equation}
where $\overrightarrow{v}$ is the absolute velocity of the particle, $v$ the magnitude of the absolute velocity and $\varphi $ a continuous
function (see [1], [3], [4] and [5]). A classical approach for $\varphi $ considered by D'Alembert, Euler and Stokes (see [5]) is:
\begin{equation}\label{*}
\varphi (v)=\lambda v^{\alpha }
\end{equation}
with $\alpha \geq 0$ a dimensionless coefficient and $\lambda >0$ the coefficient of friction. The most often used situations are: the drag
force is proportional to the velocity (Stokes' law: $\alpha =0$) and the drag force has the magnitude proportional to the square of the velocity
($\alpha =1$).

An microscopic adhesion model assumes that at the molecular scale is established an adhesion process; some particles of the resistive medium
adhere to the solid body. Adhesion is caused by intermolecular forces between the interface of materials (see [8]).

We search the answer to the question: \emph{the adhesion is the most important phenomenon related to the drag force?}

We suppose that \textit{"the drag force is produced by the adhesion of some particles on the solid body"} and others physical phenomenons
related to the drag force are neglected.

Our study focus on the evolution of the mass of the body. We suppose that the solid body is a free particle with variable mass and all particles
of the resistive medium are in a state of rest before the collision with the solid body. The equation of the mass is obtained using the next
presumption: \emph{the motion of the particle with variable mass in absence of the external force coincide with the motion of the particle with
a constant mass considering the presence of the drag force}.

\section{The evolution of the mass}

\hspace{0.5cm} In the newtonian mechanics the equation of the motion of a particle with the constant mass $m_0$ in the presence of the drag
force has the form:
\begin{equation}\label{*}
m_{0}\frac{d\overrightarrow{v}}{dt}=-\varphi (v)\overrightarrow{v}
\end{equation}

The mathematical model of the motion of a particle with variable mass is described by the generalized Meshchersky equation (see [2] and [4]):
\begin{equation}\label{*}
m\frac{d\overrightarrow{v}}{dt}=\frac{dm}{dt}(\overrightarrow{u}-\overrightarrow{v})
\end{equation}
where $m$ is the mass function of the particle and $\overrightarrow{u}$ is the absolute velocity of a medium particle before the collision with
the solid body. In our case we have $\overrightarrow{u}=\overrightarrow{0}$.

A motion of the particle is rectilinear for the both models. We study the \textbf{1-D} case. Let's consider an one-dimensional coordinates
system and at the initial moment the body is in the origin of the system and it is moving in the positive sense. We denote by $x$ the position
and $v$ is the velocity. In our case the equation (2.1) has the form:
\begin{equation}\label{*}
m_{0}\frac{dv}{dt}=-\varphi (v)v
\end{equation}
and the equation (2.2) became:
\begin{equation}\label{*}
mv=m_{0}v_{0}
\end{equation}
where $v_0$ is the initial velocity of the particle.

Using the relations (2.3) and (2.4) we obtain \emph{the time-dependent equation of mass}:
\begin{equation}\label{*}
\frac{dm}{dt}=\frac{1}{m_0}\varphi(\frac{m_0v_0}{m})m
\end{equation}
The Cauchy problem (the initial condition is $m(0)=m_0$) has the solution:
\begin{equation}\label{*}
\int_{m_0}^m \frac{1}{s\varphi (\frac{m_0 v_0}{s})}ds=\frac{t}{m_0}
\end{equation}

It is interesting to consider the dependence of the mass by position. We have \emph{the position-dependent equation of mass}:
\begin{equation}\label{*}
\frac{dm}{dx}=\frac{1}{m_0v_0}\varphi(\frac{m_0v_0}{m})m^2
\end{equation}
The solution (using the initial condition) is:
\begin{equation}\label{*}
\int_{m_0}^m \frac{1}{s^2\varphi (\frac{m_0 v_0}{s})}ds=\frac{x}{m_0v_0}
\end{equation}

\hspace{0.5cm} For the classical approaches considered in the relation (1.2) we obtain \emph{the time-dependent mass function}:
\begin{equation}\label{*}
m(t)=\left\{%
\begin{array}{ll}
    m_{0}e^{\frac{\lambda }{m_{0}}t}, & \hbox{if $\alpha=0$;} \\
    m_{0}(1+\frac{\lambda \alpha v_{0}^{\alpha }}{m_{0}}t)^{\frac{1}{\alpha }}, & \hbox{if $\alpha >0$} \\
\end{array}%
\right.
\end{equation}
and \emph{the position-dependent mass function}:
\begin{equation}\label{*}
   m(x)= \left\{%
\begin{array}{ll}
    m_{0}e^{\frac{\lambda }{m_{0}}x}, & \hbox{if $\alpha=1$;} \\
    m_{0}v_{0}(v_{0}^{1-\alpha }+\frac{\lambda (\alpha -1)}{m_{0}}x)^{\frac{1}{\alpha -1}}, & \hbox{if $\alpha \neq 1$.} \\
\end{array}%
\right.
\end{equation}

We underline the next properties of the mass:

\noindent - if the drag force verifies the Stokes' law then the time-dependent mass function of the particle is not depending on the initial
velocity;

\noindent - if the magnitude of the drag force is proportional to the square of the velocity then the position-dependent mass function is not
depending on the initial velocity;

\noindent - for $\varphi $ of the form (1.2) and an initial velocity $v_{0}\neq 0 $ we have $\lim\limits_{t\rightarrow \infty }m(t)=\infty .$

\section{Application: the Table Tennis ball mass}

\hspace{0.5cm} To decide if the adhesion is the most important phenomenon related to the drag force we study the evolution of a Table Tennis
ball in our model.

A Table Tennis ball has a spherical form with a radius $r=2$ cm and an initial mass $m_{0}=0.0027$ kg. The adherence of some particles of
resistive medium to the Table Tennis ball induce that the form of the ball, at a moment $t>0,$ is not necessary a sphere and the radius $r$ is
not constant. In our model we neglect this aspect and we suppose that during the motion the Table Tennis ball maintains its spherical shape of
constant radius.

We consider two relevant cases. \vspace{2mm}

\noindent \textbf{I. The mass using the Stokes' law}

We suppose that the resistive medium has a coefficient of viscosity $\eta .$ According to the Stokes' law we have a drag force of the form:
\begin{equation}\label{*}
F_{f}=-\lambda v
\end{equation}
The coefficient of friction is (see [6]) $\lambda =6\pi \eta r$ and we consider that the resistive medium is the water at the temperature
$20^{o}$ $C$. In this case the coefficient of viscosity is $\eta =1.005$ $N\,s\, m^{-3}$ and the coefficient of friction is $\lambda =0.000378$
$kg/s.$ Using the relations (2.9) and (2.10) we obtain:
\begin{equation}\label{*}
    \left\{%
\begin{array}{ll}
    m(t)=0.0027e^{0.14\cdot t} \\
    m(x)=\frac{1}{370.37-51.97\frac{x}{v_{0}}} \\
\end{array}%
\right.
\end{equation}

We observe that the mass of the Table Tennis ball doubles in a time period $t^{*}=4.93$ seconds. \vspace{3mm}
\newpage

\noindent \textbf{II. The mass using a drag force proportional to the square of the velocity}

The drag force has the form:
\begin{equation}\label{*}
F_{f}=-\lambda v^2
\end{equation}
The coefficient of friction is $\lambda =0.87r^{2}$ (see [10]) and we consider that the resistive medium is equivalent the air. In this case the
coefficient has the value $\lambda =0.000348\,kg/m$. We obtain:
\begin{equation}\label{*}
    \left\{%
\begin{array}{ll}
    m(t)=0.0027+(0.000348\cdot v_{0})\cdot t \\
    m(x)=0.0027e^{0.128\cdot x} \\
\end{array}%
\right.
\end{equation}
We observe that the Table Tennis ball mass doubles after a distance of $l^{**}=5.37$ meters.

\section{Conclusion}

\hspace{0.5cm} This paper is presenting a model which considers the influence of a resistive medium in terms of the phenomenon of macroscopic
adherence. In general case the evolution of the solid body mass is calculated by the relations (2.6) and (2.8) and in the classical approaches
of the drag force (see relation (1.2)) by the relations (2.9) and (2.10).

The simulation of the mass function of a Table Tennis ball reveals that \emph{the adhesion phenomenon can't explain alone the drag force}.

\subsection*{Acknowledgement}

\hspace{0.5cm} The author has been supported by the Romanian Ministry of Education and Research, Grant CNCSIS 95GR 2007/2008.

\subsection*{References}

[1] \textit{Com\u{a}nescu D.}, 2007, Mathematical Methods in Mechanics, Ed. Mirton.

\noindent [2] \textit{Com\u{a}nescu D.}, 2005, Rectilinear Motions of a Rocket with Variable Mass in a Force Field Created by Two Fixed Bodies,
Proc. 4-th Int. Conf. Nonlin. Prob. in Aviation and Aerospace ICNPAA, European Conference Publications, Cambridge, 2005, 221-228.

\noindent [3] \textit{Com\u{a}nescu D.}, 2004, Models and Methods in Material Point Mechanics, Ed. Mirton.

\noindent [4] \textit{Drago\c{s} L.}, 1976, Principles of Analitical Mechanics, Ed. Tehnica, Bucuresti.

\noindent [5] \textit{Iacob C.}, 1980, Theoretical Mechanics, Ed. Didactica si Pedagogica, Bucuresti.

\noindent [6] \textit{Sears F.W., Zemansky M.W., Young H.D.}, 1976, University Physics, Fifth Edition, Addison-Wesley Pub. Comp., 1976.

\noindent [7] \textit{Starzhinskii V.M.}, 1982, An Advanced Course of Theoretical Mechanics for Engineering Students, MIR Pub., Moscow, 1982.

\noindent [8] \textit{Wautelet M. et coll., }2003, Les Nanotechnologies, Dunod, Paris.

\noindent [9] http://www.nano-world.org/friction module/content/

\noindent [10] http://scott-yost.bayler.edu/phy1422f05/

\end{document}